\documentclass[
reprint,
superscriptaddress,
 amsmath,amssymb,
 aps,
floatfix
]{revtex4-1}

\usepackage{longtable}
\usepackage{makecell}
\usepackage{amsmath,etoolbox}
\usepackage{amsthm}
\usepackage{amssymb}
\usepackage[left=2.5cm,right=2.5cm,top=3cm,bottom=3cm,bindingoffset=0cm]{geometry}
\usepackage{dsfont}
\usepackage{bm}
\usepackage{leftidx}
\usepackage{float}
\usepackage[makeroom]{cancel}
\usepackage{indentfirst}
\usepackage{hyperref}
\usepackage{tikz}
\usetikzlibrary{quantikz}
\usepackage{upgreek}
\usepackage[caption=false]{subfig}
\usepackage{graphicx}
\usepackage{tikz-cd}
\usepackage[titletoc,title]{appendix}

\usepackage{adjustbox}
\usepackage{mathtools}
\usepackage{multirow}
\usepackage{xargs}
\usepackage{xstring}
\usepackage{rotating}

\usetikzlibrary{matrix,arrows,decorations.pathmorphing}
\newenvironment{sloppypar*}{\sloppy\ignorespaces}{\par}
\allowdisplaybreaks 

\newcommand {\sket} [1] {| #1 \rangle}

\newcommand {\sand} [3] {\langle #1 | #2 | #3 \rangle}

\newcommand{\bma} {\begin{pmatrix}}
\newcommand{\ema} {\end{pmatrix}}

\renewcommand{\d}[1]{\ensuremath{\operatorname{d}\!{#1}}}

\newcommandx*\hilbdimK[1][1=]{\mathcal{D}_{K\IfEqCase{#1}{{}{}}[=#1]}}
\newcommandx*\hilbdimKQ[2][1=,2=]{\mathcal{D}_{K\IfEqCase{#1}{{}{}}[=#1],Q\IfEqCase{#2}{{}{}}[=#2]}}

\newcommand*{\vv}[1]{\vec{\mkern0mu#1}}

\newcommandx*\bound[3][,2=,3=]{#1_{K
\IfEqCase{#2}{{}{}}[,\,#2]}^{
\IfEqCase{#3}{{}{}}[(#3)]
}}

\DeclareMathAlphabet{\mathdutchcal}{U}{dutchcal}{m}{n}

\def\ket#1{\left\vert #1 \right\rangle}
\def\bra#1{\left\langle #1 \right\vert}

\newcommand{\be}{\begin{equation}}
\newcommand{\ee}{\end{equation}}
\newcommand{\bp}{\begin{pmatrix}}
\newcommand{\ep}{\end{pmatrix}}
\newcommand{\ben}{\begin{enumerate}}
\newcommand{\een}{\end{enumerate}}

\def\mystrut(#1,#2){\vrule height #1pt depth #2pt width 0pt}

\let\oldFootnote\footnote
\newcommand\nextToken\relax
\renewcommand\footnote[1]{\oldFootnote{#1}\futurelet\nextToken\isFootnote}
\newcommand\isFootnote{\ifx\footnote\nextToken\textsuperscript{,}\fi}

\usepackage[shortcuts]{extdash}

\newcommand*\standardbin{+}

\makeatletter

\newcommand{\doublewidetilde}[1]{{%
  \mathpalette\double@widetilde{#1}%
}}
\newcommand{\double@widetilde}[2]{%
  \sbox\z@{$\m@th#1\widetilde{#2}$}%
  \ht\z@=.9\ht\z@
  \widetilde{\box\z@}%
}

\patchcmd{\env@cases}
  {\quad}
  {{ \ }}
  {}{}

\newcommand*\tabularbin[1]{%
  \mathbin{\mathpalette{\@tabularsym\standardbin}{#1}}%
}

\newcommand*\@tabularsym[3]{%
  \setbox\z@\hbox{$#2#1\m@th$}%
  \hbox to\wd\z@{\hss$#2#3\m@th$\hss}%
}

\newcommand*{\myrulefill}[3][]{%
  \makebox[#2]{#1%
    \leaders\hrule height \dimexpr.5ex+.2pt\relax depth \dimexpr -.5ex+.2pt\relax \hfill
    \enskip{#3}\enskip
    \leaders\hrule height \dimexpr.5ex+.2pt\relax depth \dimexpr -.5ex+.2pt\relax \hfill\kern0pt}
}

\usetikzlibrary{arrows,positioning}
\tikzset{
    >=stealth',
    punkt/.style={
           rectangle,
           rounded corners,
           draw=black, very thick,
           text width=6.5em,
           minimum height=2em,
           text centered},
    pil/.style={
           ->,
           thick,
           shorten <=2pt,
           shorten >=2pt,}
}

\makeatother

\begin{document}

\title{Light-Front Field Theory on Current Quantum Computers}

\author{Michael Kreshchuk}
\affiliation{Department of Physics and Astronomy, Tufts University, Medford, MA 02155, USA}
\author{Shaoyang Jia}
\affiliation{Department of Physics and Astronomy, Iowa State University, Ames, IA 50011, USA}
\author{William M. Kirby}
\affiliation{Department of Physics and Astronomy, Tufts University, Medford, MA 02155, USA}
\author{Gary Goldstein}
\affiliation{Department of Physics and Astronomy, Tufts University, Medford, MA 02155, USA}
\author{James P. Vary}
\affiliation{Department of Physics and Astronomy, Iowa State University, Ames, IA 50011, USA}
\author{Peter J. Love}
\affiliation{Department of Physics and Astronomy, Tufts University, Medford, MA 02155, USA}
\affiliation{Computational Science Initiative, Brookhaven National Laboratory, Upton, NY 11973, USA }

\begin{abstract}

We present a quantum algorithm for simulation of quantum field theory in the light-front formulation and demonstrate how existing quantum devices can be used to study the structure of bound states in relativistic nuclear physics.
Specifically, we apply the Variational Quantum Eigensolver algorithm to find the ground state of the light-front Hamiltonian
obtained within the Basis Light-Front Quantization framework.
As a demonstration, we calculate the mass, mass radius, decay constant, electromagnetic form factor, and charge radius of the pion on the \texttt{ibmq\_vigo} chip.
We consider two implementations based on different encodings of physical states, and propose a development that may lead to quantum advantage.
This is the first time that the light-front approach to quantum field theory has been used to enable simulation of a real physical system on a quantum computer.

\end{abstract}

\maketitle

\section{Introduction\label{sec:intro}}

The light-front quantization framework of quantum field theories (QFTs) is well-adapted for digital quantum simulation.
We demonstrated this in our previous work by developing quantum algorithms based on simulating time evolution and adiabatic state preparation~\cite{Kreshchuk:2020dla}.
In the present paper we aim for near-term devices by showing how to formulate the relativistic bound state problem as an instance of the Variational Quantum Eigensolver (VQE) algorithm~\cite{peruzzo2014variational,mcclean2016theory,subspace,discriminative,excited}.
VQE is a hybrid quantum-classical algorithm for finding low-lying eigenvalues and eigenstates of a given Hamiltonian, which can be implemented on existing quantum computers.
We are thus able to run example simulations on $\texttt{ibmq\_vigo}$, one of IBM's publicly available quantum processors.

For an efficient Hamiltonian formulation of quantum field theory, we use the framework of Basis Light-Front Quantization (BLFQ)~\cite{varybasis,Zhao:2014hpa} and choose a basis tailored to the symmetries and dynamics specific to a particular physical system. Having much in common with \emph{ab initio} methods in quantum chemistry and nuclear theory, it serves as an ideal framework for testing near-term devices by solving problems such as calculation of hadronic spectra~\cite{LI2016118,Li:2017mlw,Tang:2018myz,Tang:2019gvn} and parton distribution functions~\cite{Lan:2019vui,Lan:2019img,Lan:2019rba}.

\begin{figure*}
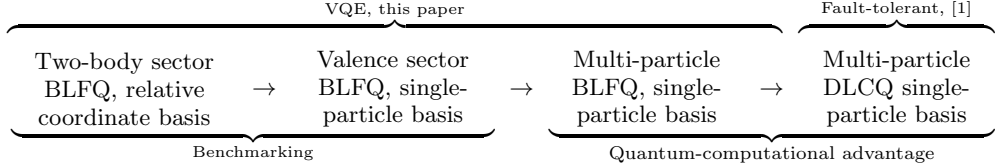

\centering
\begin{equation*}
    \lefteqn{\overbrace{\mystrut(22.58,0)\hspace{10.3cm}}^{\text{VQE, this paper}}}
    \underbrace{
    \parbox{3.0cm}{Two-body sector\\BLFQ, relative\\coordinate basis}
    \ \rightarrow \
    \parbox{2.7cm}{Valence sector\\BLFQ, single-\\particle basis}}_{\text{Benchmarking}}
    \ \rightarrow \
    \underbrace{
    \parbox{2.6cm}{Multi-particle\\BLFQ, single-\\particle basis}
    \ \rightarrow \
    \overbrace{\mystrut(22.58,0)\parbox{2.6cm}{Multi-particle\\DLCQ single-\\particle basis}}^{\text{Fault-tolerant, \cite{Kreshchuk:2020dla}}}
    }_{\text{\scriptsize \centering Quantum-computational advantage}}
\end{equation*}
\caption{
Flow of increasing complexity and computational resources (left to right) for quantum simulation of quantum field theory in the light-front formulation.
\label{fig:hierarchy}
}
\end{figure*}

Within BLFQ, a field is expanded in terms of second-quantized Fock states representing occupancies of modes (first-quantized basis functions), and there is no \emph{a priori} limit on the degrees of freedom~\cite{varybasis}. Accordingly, our algorithms are designed to efficiently simulate
QFT applications where particle number is not conserved.
However, for QFTs at low resolution or for phenomenological applications, BLFQ is often restricted to the valence degrees of freedom, so we adopt this restriction in order to implement quantum simulations on an existing quantum chip.
These experiments represent the first stage shown in Fig.~\ref{fig:hierarchy}, which illustrates a progression of methods that scale towards fault-tolerant simulation of QFTs in the quantum supremacy regime.
However, the methods we propose apply to the first three stages in Fig.~\ref{fig:hierarchy} (the final stage was discussed in~\cite{Kreshchuk:2020dla}).

For our experimental demonstration we consider the dynamics of valence quarks for light mesons on the light front using the Hamiltonian form~\cite{basislightmesons}. This Hamiltonian includes the kinetic energy, the confinement potential in both the longitudinal and the transverse directions~\cite{Li:2017mlw}, and the Nambu--Jona-Lasinio (NJL) interaction~\cite{Klevansky:1992qe} to account for the chiral interactions among quarks.
The dependence of the light-front wave functions for valence quarks on the relative momentum is expanded in terms of the adopted modes, which are orthonormal basis functions. After imposing finite cut-offs in this expansion, the light-front Hamiltonian becomes a Hermitian matrix in the resulting basis representation. We use the same scheme as in~\cite{basislightmesons} to fix our model parameters at each choice of basis cut-offs.

We implement VQE for this model on the \texttt{ibmq\_vigo} processor. We minimize the mass-squared of a pion obtained from a variational ansatz for its wavefunction. Using the resulting ansatz, we compute the decay constant, mass radius, and elastic form factor of the pion on the quantum processor. We thus demonstrate that the light-front formulation of QFT enables calculations of properties of composite particles in relativistic field theories on existing quantum processors.

\section{Basis Light-Front Quantization\label{sec:blfq}}

In order to solve for the light front wave functions of the valence quarks inside light mesons, we use the effective Hamiltonian of the form~\cite{basislightmesons}
\begin{equation}\label{eq:decompose_Heff}
    H_{\mathrm{eff}}=H_0+H^{\mathrm{eff}}_{\mathrm{int}} \ .
\end{equation}
In \eqref{eq:decompose_Heff}, the term $H_0$ contains the kinetic energy of the valence quarks, the transverse confinement potential, and the longitudinal confinement potential. For a meson carrying light-front momentum $P^+$ and transverse momentum $\overrightarrow{P}^\perp$, the valence Fock sector Hamiltonian $H_0$ takes the form
\begin{align}\label{eq:H0_def}
\begin{alignedat}{8}
&H_0 =\dfrac{(\overrightarrow{\kappa}^\perp)^2+\mathbf{m}^2}{x}+\dfrac{(\overrightarrow{\kappa}^\perp)^2+\overline{\mathbf{m}}^2}{1-x}
\\&+b^4x(1-x)\overrightarrow{r}_\perp^2
    -\dfrac{b^4}{(\mathbf{m}+\overline{\mathbf{m}})^2}\partial_xx(1-x)\partial_x \ ,
\end{alignedat}
\end{align}
\begin{sloppypar}
where $x=k^+/P^+$ is the longitudinal momentum fraction carried by the valence quark, $b$ specifies the strength of the confinement potentials, ${\overrightarrow{\kappa}^\perp=\overrightarrow{k}^\perp-x\overrightarrow{P}^\perp}$ is the relative transverse momentum of the valence quarks, and $\overrightarrow{r}_\perp$ is the conjugate variable of $\overrightarrow{\kappa}^\perp$. $k^\mu$
are the 4-momentum components of the valence quark. The masses of the valence quark and valence antiquark are $\mathbf{m}$ and $\overline{\mathbf{m}}$, respectively.
The remaining part of the strong interaction between quarks, $H^{\mathrm{eff}}_{\mathrm{int}}$, is modeled using the scalar-pseudoscalar channel of the color-singlet NJL model~\cite{Klevansky:1992qe}:
\end{sloppypar}
\begin{align}\label{eq:H_eff_NJL_pi_ori}
\begin{alignedat}{8}
H^{\mathrm{eff}}_{\mathrm{int}} = H_{\mathrm{NJL},\pi}^{\mathrm{eff}}
    =&\int \d{x}^-\int\d{}\overrightarrow{x}^\perp\,\bigl(-\dfrac{G_\pi P^+}{2}\bigr)
    \\&\times\left[\left(\overline{\psi}\psi \right)^2+\left(\overline{\psi}i\gamma_5\overrightarrow{\tau}\psi \right)^2 \right] \ .
\end{alignedat}
\end{align}
Here $x^-$ and $\overrightarrow{x}^\perp$ are the single-particle light-front coordinates, $\psi$ is the fermion field operator and $G_\pi$ is the NJL coupling constant.
We then expand \eqref{eq:H_eff_NJL_pi_ori} into the appropriate combinations of ladder operators for the quark fields.

Within the BLFQ, the light-front wave functions of the valence quarks are expressed as~\cite{basislightmesons}
\begin{equation}
\label{eq:Psi_meson_qqbar}
\begin{alignedat}{9}
&\ket{\Psi(P^+,\overrightarrow{P}^\perp)}
=\sum_{r,s}\int_{0}^{1}\dfrac{dx}{4\pi x(1-x)}
\\&
\hspace{-.05cm}\times\int\dfrac{d\overrightarrow{\kappa}^\perp}{(2\pi)^2}\,\psi_{rs}(x,\overrightarrow{\kappa}^\perp)
\times b_r^\dagger(xP^+,\overrightarrow{\kappa}^\perp+x\overrightarrow{P}^\perp)
\\& \hspace{.81cm}
\times d_s^\dagger((1-x)P^+,-\overrightarrow{\kappa}^\perp+(1-x)\overrightarrow{P}^\perp)\sket{0} \ .
\end{alignedat}
\end{equation}
\begin{sloppypar}
The ladder operators $b^\dagger_r$ and $d^\dagger_r$ create a quark and an antiquark of spin $r$ from the light-front vacuum, and obey the usual anticommutation relations ${\{b_r,b^\dagger_s\}=\{d_r,d^\dagger_s\}=\delta_{rs}}$ (all other anticommutators being zero).
The light-front wave function for the valence quarks is then expanded in the following orthonormal basis:
\end{sloppypar}
\begin{equation}\label{eq:psi_rs_basis_expansions}
\begin{alignedat}{9}
\psi_{rs}(x,\overrightarrow{\kappa}^\perp)
 &= \sum_{nml} \psi(n,m,l,r,s)
\\&
\times\phi_{nm}\left(\dfrac{\overrightarrow{\kappa}^\perp}{\sqrt{x(1-x)}};b\right)
\chi_l(x)
\ ,
\end{alignedat}
\end{equation}
where $\phi_{nm}$ is a 2-dimensional (2D) harmonic oscillator eigenfunction, $\chi_l$ is the longitudinal basis function related to Jacobi polynomials~\cite{basislightmesons}, and $n$, $m$, and $l$ are the radial, angular, and longitudinal basis quantum numbers respectively.
The momentum scale of the harmonic oscillator function is chosen identical to the confinement strength in Eq.~\eqref{eq:H0_def}. In the representation in which analytic expressions exist for these basis functions, $H_{\rm 0}$ is diagonal. Furthermore, the matrix elements of the full Hamiltonian~\eqref{eq:decompose_Heff} in this representation can be calculated analytically~\cite{basislightmesons}.

For the experimental demonstrations below, we use the light meson BLFQ Hamiltonian with the minimal choice of basis function cutoffs and model parameters specified in Tab.~\ref{tab:model_parameters}. In the zero azimuthal angular momentum block, the Hamiltonian describes the interaction of quarks whose momentum-space wave function is in
the lowest eigenstate of $H_0$:
\begin{table}[]
    \hspace{-.79cm}
    \begin{tabular}{ccc}
	\hline
	$\mathbf{m}$ & $\kappa$ & $G_\pi$ \\
	\hline
	$337.01~\mathrm{MeV}$ & $227.00~\mathrm{MeV}$ & $250.785~\mathrm{GeV}^{-2}$ \\
	\hline
    \end{tabular}
    \caption{Model parameters of the BLFQ-NJL model. The antiquark mass is identical to the quark mass.}
    \label{tab:model_parameters}
\end{table}
\begin{equation}
\label{eq:ham_ex}
\hspace{-.2cm}
H=
 \begin{pmatrix}
 640323 & 139872 & -139872 & -107450 \\
 139872 & 346707 &  174794 &  139872 \\
 -139872 & 174794 & 346707 & -139872 \\
 -107450 & 139872 & -139872 & 640323
 \end{pmatrix} \\ ,
\end{equation}
where the matrix elements are in units of $\mathrm{MeV}^2$. The size of $H$ reflects the $4$ possible spin configurations of the valence quarks. In this case the NJL interaction takes the role of the spin-orbit interaction of quarks.
The lowest eigenvalue of $H$ corresponds the squared mass of the pion, ${m_\pi^2 = \left(139.6~\text{MeV}\right)^2}$.
Note that in the light-front formulation the Hamiltonian is the invariant mass-squared operator~\cite{brodsky98a}.

\section{Variational Quantum Eigensolver}
\begin{sloppypar}
VQE is an approach to finding Hamiltonian eigenvalues, in which a quantum processor is used as part of a hybrid quantum-classical algorithm~\cite{peruzzo2014variational}.
In VQE, a quantum computer is used to evaluate the Hamiltonian expectation value for a given variational state, while a classical computer performs a gradient search to minimize the expectation value. In order to formulate a physical problem as a VQE instance, one has to a)~Establish a correspondence between the physical states and the multi-qubit states of a quantum computer, b)~Prepare a parametrized ansatz state on the quantum computer ${\sket{\psi(\vv{\theta})} = U(\vv{\theta}) \sket{\psi_0}}$ (${\sket{\psi_0}}$ is some easy to prepare reference state), c)~Evaluate the Hamiltonian expectation value ${E(\vv{\theta}) =
\sand{\psi(\vv{\theta})}{\widehat{H}}{\psi(\vv{\theta})}}$ 
by sampling on the quantum computer, d)~Send the estimated value $E(\vv{\theta})$ to the classical optimizer to determine the set of parameters for the next iteration of the algorithm.
\end{sloppypar}

\begin{figure*}
\centering
\subfloat[]{
\adjustbox{valign=b}{
\begin{tikzpicture}
\node[scale=1]{
\begin{quantikz}[column sep = .2cm, row sep = .1cm]
\lstick[wires=4]{$\sket{0}^{\otimes 4}$} & \qw & \qw & \qw & \qw & \gate{R_y(\theta_2)} & \ctrl{1} & \qw
\\
& \qw & \gate{X} & \ctrl{1} & \gate{X} & \ctrl{-1} & \gate{X} & \qw
\\
& \qw & \qw & \gate{R_y(\theta_1)} & \ctrl{-1} & \ctrl{1} & \gate{X} & \qw
\\
& \qw & \qw & \qw & \qw & \gate{R_y(\theta_3)} & \ctrl{-1} & \qw
\end{quantikz}
};
\end{tikzpicture}
}
\label{fig:circuitdir}
}
\qquad
\hfill\subfloat[]{
\adjustbox{ valign = b}{
\begin{tikzpicture}
\node[scale=1]{
\begin{quantikz}[column sep = .2cm, row sep = .1cm]
\lstick[wires=2]{$\sket{0}^{\otimes 2}$} & \gate{R(\theta_1,\phi_1,\lambda_1)} & \gate{X} & \gate{R(\theta_3,\phi_3,\lambda_3)} & \qw
\\
& \gate{R(\theta_2,\phi_2,\lambda_2)} & \ctrl{-1} & \qw  & \qw \\[.55cm]
\end{quantikz}
};
\end{tikzpicture}
}
}

\label{fig:circuitbin}
\caption{Ansatz circuits for preparing an arbitrary superposition of single-particle Fock states with real coefficients in the direct encoding (a) and
compact encoding (b). $R_y(\theta)$ denotes a single-qubit rotation through an angle $\theta$ about the $y$-axis. $R$ is an arbitrary single-qubit rotation~\cite{Qiskit}. Gate parameters are obtained by means of the Qiskit \texttt{initialize} routine, as in~\cite{synth}.}
\label{fig:circuit}
\end{figure*}
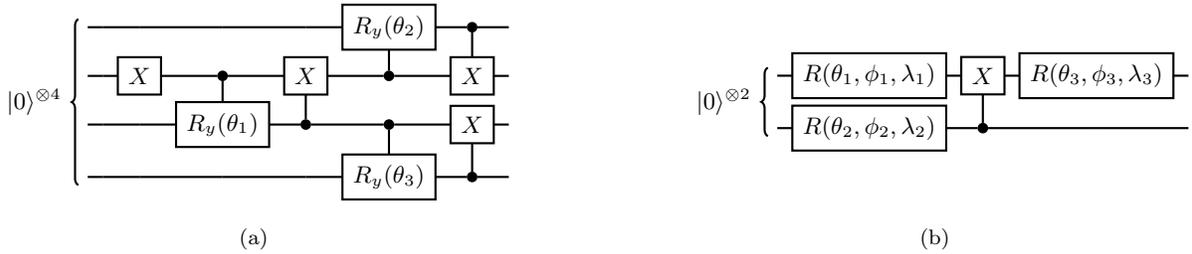

We explore two approaches to simulation in the BLFQ formulation. The first uses \emph{direct encoding} of Fock states in qubit states, meaning that the occupation of each mode is represented in a fixed register of qubits.
Since we are using the relative momentum basis and working within the valence sector of the Fock space, the basis Fock states
only contain one occupied mode.
We employ the Jordan-Wigner (JW) encoding~\cite{jordanwigner}, which is commonly used in quantum chemistry~\cite{aspuru2005simulated,somma2002simulating}, and in our case simply means encoding the occupation of each mode in a single qubit.
Any superposition of such encoded states can be prepared using the simple circuit given in Fig.~\ref{fig:circuitdir}.
For multi-particle states, one could switch to the more efficient Bravyi-Kitaev encoding~\cite{bravyi2002fermionic,BK2015}, and use the Unitary Coupled Cluster ansatz~\cite{Romero_2018}.

A different approach is based on \emph{compact encoding}~\cite{aspuru2005simulated,Kreshchuk:2020dla}, in which only the quantum numbers of occupied modes are stored; in our case this amounts to storing the index of the single occupied mode in binary form.
Since the number of qubits required for storing a single-particle Fock state is logarithmic in the number of modes, one can use arbitrary state preparation as an ansatz circuit, given in Fig.~\ref{fig:circuitbin}.

The expectation value of the Hamiltonian is calculated as
\begin{equation}
    \label{eq:HPauliexp}
    \sand{\psi(\vv{\theta})}{\widehat{H}}{\psi(\vv{\theta})}
    = \sum_i h_i \sand{\psi(\vv{\theta})}{P_i}{\psi(\vv{\theta})} \ .
\end{equation}
The expectation values of the individual Pauli terms on the RHS of~\eqref{eq:HPauliexp} can be efficiently measured via sampling from the state $\sket{\psi(\vv{\theta})}$~\cite{peruzzo2014variational}.

\begin{figure}
\centering
\hspace{-.4cm}
\includegraphics[width = .48\textwidth]{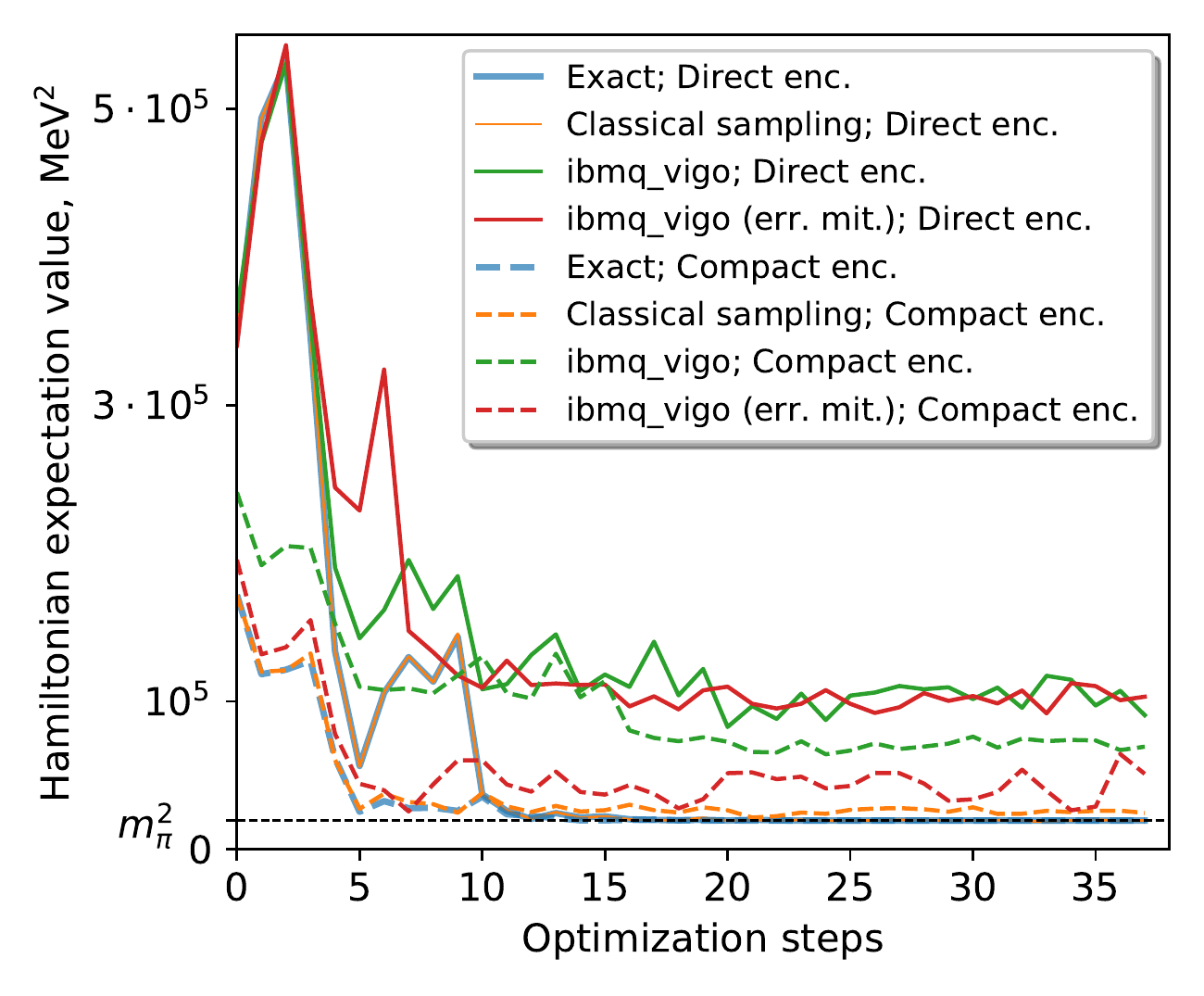}
\caption{
Results of the VQE minimization algorithm in the compact and direct encodings. Each point was obtained from 8192 samples per term on \texttt{ibmq\_vigo} chip. Note that here $m_\pi^2=(139.6\text{MeV})^2$ is the lowest eigenvalue of the Hamiltonian, by definition (see \eqref{eq:ham_ex} and the associated discussion).}
\label{fig:vqe_minimization}
\end{figure}

\section{Results}
\begin{figure}
\hspace{-1.cm}
\includegraphics[width = .49\textwidth]{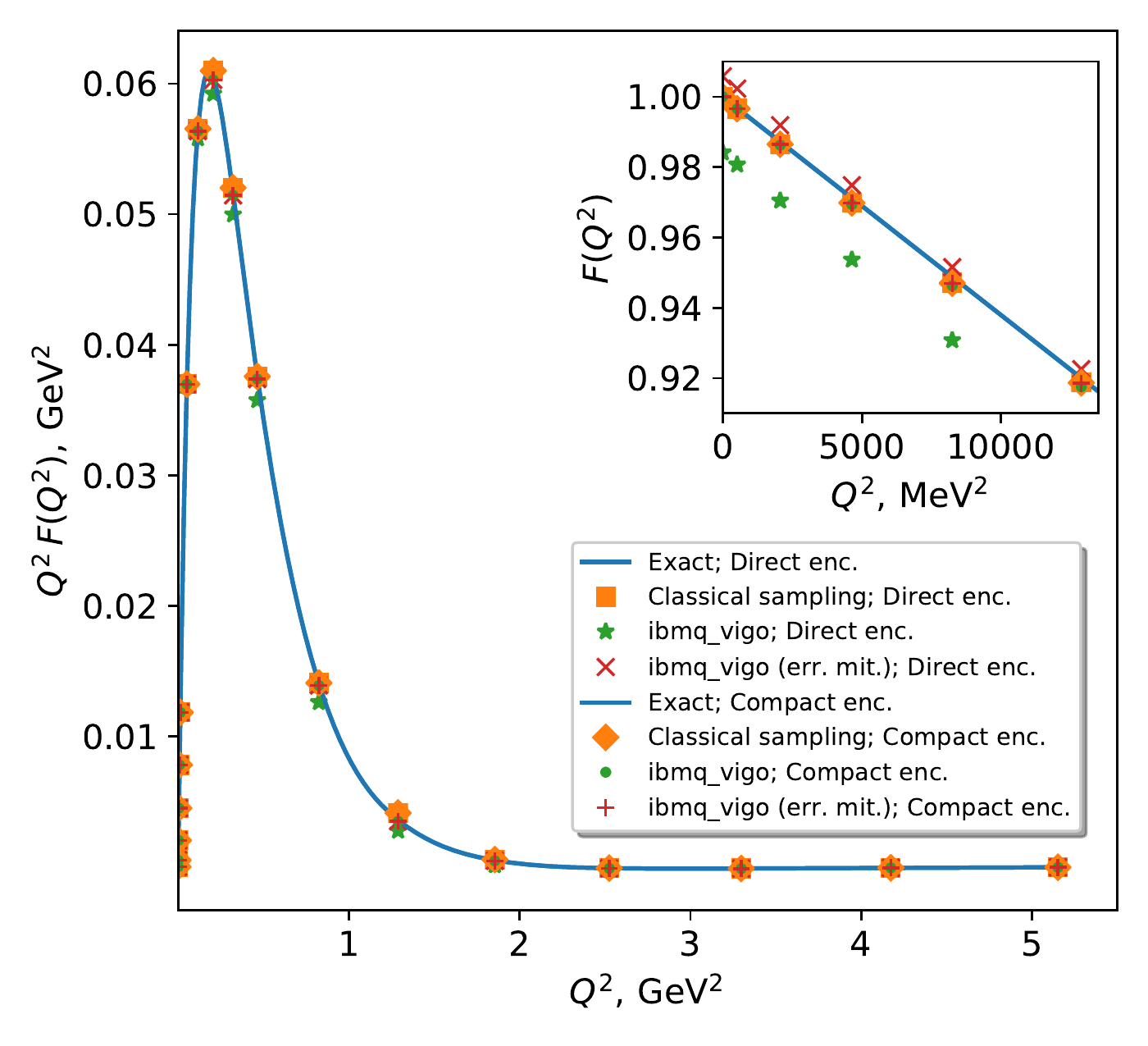}
\caption{Pion elastic form factor, obtained from 8192 samples per term on the \texttt{ibmq\_vigo} processor, with and without measurement error mitigation. [SJ: the vertical axis should have the unit of $\mathrm{GeV}^2$ for the main plot (not in the magnification).]}
\label{fig:form_factor}
\end{figure}

We implemented VQE on the \texttt{ibmq\_vigo} quantum processor using both direct and compact encodings with and without error mitigation provided by Qiskit~\cite{Qiskit}.
In Fig.~\ref{fig:vqe_minimization}, we show the experimentally obtained energies at each minimization step, as well as the exact values and those obtained by classical sampling from the exact probability distributions (the latter illustrates the performance of a noiseless quantum computer).
The improvement due to measurement error mitigation was significant only for the compact encoding, and led to the best convergence to the true ground state energy out of the experimental methods.

We evaluated additional observables in the ground state.
In Tab.~\ref{tab:results}, we show the accuracies obtained using each technique: exact evaluation, classical sampling, and sampling on the \texttt{ibmq\_vigo} chip with and without measurement error mitigation.
We prepared the ground state on the \texttt{ibmq\_vigo} chip by using the parameters obtained in our VQE minimization.
The observables we measured are pion mass, mass radius, and decay constant.
As expected, the results obtained using the compact encoding are consistently more accurate than those obtained using the direct encoding, since the corresponding ansatz circuits are shorter.
Measurement error mitigation consistently improves the accuracies in the compact encoding, and provides no benefit in the direct encoding.
However, we do see that in nearly all cases the quantum methods are approximately correct, with the method using compact encoding and measurement error mitigation approaching the performance of classical sampling.

\begin{table*}[]
\begin{tabular}{|l|c|c|c|c|c|c|}
\hline
 & \multicolumn{3}{c|}{Direct encoding} & \multicolumn{3}{c|}{Compact encoding} \\ \hline
 & \makecell{Classical\\sampling}        &  \texttt{ibmq\_vigo}     & \makecell{\texttt{ibmq\_vigo},\\err. mit.} &  \makecell{Classical\\sampling}  &     \texttt{ibmq\_vigo}      & \makecell{\texttt{ibmq\_vigo},\\err. mit.}          \\ \hline
$m_\pi^2$, no constant  &  0.48\%    &    7.6\%  & 7.5\%  &    0.01\%       &     11.6\%      &    6.2\%       \\ \hline
$m_\pi^2$&    0.90\%     & 14.1\% & 14.0\%   &     0.08\%      &  12.7\%         &     9.1\%      \\ \hline
$\langle r_{m} \rangle^2$, no constant &   0.45\%        &6.6\%       &  7.2\% & 0.43\%   &   29.4\%        &      7.1\%     \\ \hline
$\langle r_{m} \rangle^2$ &  0.65\%       & 9.5\% & 10.4\%  &    0.01\%       &      6.4\%     &    1.6\%       \\ \hline
$f_\pi$, no constant &    0.05\%       &  59.8\%      &  59.0\% & 0.21\%  &  29.2\%         &  7.6\%         \\ \hline
$f_\pi$  &    0.02\%      & 21.0\% &  20.7\% &    0.14\%       &      13.0\%     &     5.1\%      \\ \hline
\end{tabular}
\caption{Fractional errors, expressed as percentages, in estimates of various observables calculated in the exact ground state. The observables are pion mass squared ($m_\pi^2$), mass radius squared ($\langle r_m\rangle^2$), decay constant ($f_\pi$), and charge radius ($\langle r_{\rm c} \rangle^2$).
These were obtained from 8192 samples per term on \texttt{ibmq\_vigo} chip, with and without measurement error mitigation. Classical sampling means sampling from the exact probability distribution. Observables are shown both including constant terms (the physically relevant values), and not including them (the measured values). For $m_\pi^2$, the exact $m_\rho^2$ is used for normalization.
}
\label{tab:results}
\end{table*}

\begin{sloppypar}
We computed the pion elastic form factor~$F(Q^2)$, obtaining the results shown in Fig.~\ref{fig:form_factor}.
Based on these data, we computed the pion charge radius as ${\langle r_{\mathrm{c}}^2\rangle =-6\d{F}(Q^2)/\d{Q}^2|_{Q^2=0}}$. The values obtained using the quantum computer match those obtained via the state vector representation, $\sqrt{{r_{\mathrm{c}}^2}} = 1.24~\mathrm{fm}$, within a few percent precision.
These calculations illustrate that our algorithm provides reasonable results for physically meaningful quantities even with the noisy and limited quantum resources that are currently available.
\end{sloppypar}

\section{Discussion}

In this work, we demonstrated how one can use existing quantum processors to perform calculations in relativistic field theories in the light-front formulation.
The methods we proposed apply to the multi-particle setting, which can potentially reach the regime of quantum advantage.
While designing a scalable VQE ansatz for the compact encoding remains an open problem, using the direct encoding allows one to readily employ techniques developed for digital simulation of quantum chemistry.
We have thus demonstrated the viability of quantum simulation in the light-front formulation, using methods that can be scaled to exploit the available quantum resources, from existing noisy intermediate-scale quantum machines up to the crossover into fault-tolerance.

\begin{acknowledgements}
W.~M.~K. acknowledges support from the National Science Foundation, Grant No. DGE-1842474.
P.~J.~L. acknowledges support from the National Science Foundation, Grant No. PHY-1720395, and from Google Inc.
M.~K. and G.~G.~acknowledge support from DOE HEP Grant No. DE-SC0019452. S.~J. and J.~P.~V. acknowledge support from DOE Grant Nos. DE-FG02-87ER40371 and DE-SC0018223.
This work was supported by the NSF STAQ project (PHY-1818914).
\end{acknowledgements}

\bibliography{references}

\end{document}